\documentclass[
 reprint,
 nofootinbib,
 showkeys,
 amsmath,
 amssymb,
 aps,
 pra,
 floatfix,
 superscriptaddress,
 longbibliography
 ]{revtex4-2}

\usepackage{graphicx}
\usepackage{multirow}
\usepackage[dvipsnames]{xcolor}
\usepackage{dcolumn}
\usepackage{bm}
\usepackage{dsfont}
\usepackage{physics}
\usepackage[hidelinks, colorlinks=false, pdfborder={0 0 0}]{hyperref}

\usepackage{ulem}

\usepackage[version=3]{mhchem} 
\usepackage[dvipsnames]{xcolor}
\usepackage{textcomp}
\usepackage{graphicx}
\usepackage{amsmath}
\usepackage{fixmath}
\usepackage{amsfonts}
\usepackage{amssymb}
\usepackage{upgreek}
\usepackage{bm}
\usepackage{ulem}
\usepackage{siunitx}

\newcommand{\micron}{\hbox{\textmu}\text{m}}

\begin{document}

\title{Magnetic switching of self-hybridized exciton-polaritons in CrSBr photonic crystal slabs}

\author{Tatiana D. Gorelkina}
\affiliation{School of Physics and Engineering, ITMO University, Saint Petersburg 197101, Russia}
\author{Ivan E. Kalantaevskii}
\affiliation{School of Physics and Engineering, ITMO University, Saint Petersburg 197101, Russia}
\author{Artem N. Abramov}
\affiliation{School of Physics and Engineering, ITMO University, Saint Petersburg 197101, Russia}
\author{Ksenia A. Gasnikova}
\affiliation{School of Physics and Engineering, ITMO University, Saint Petersburg 197101, Russia}
\author{Prokhor A. Alekseev}
\affiliation{School of Physics and Engineering, ITMO University, Saint Petersburg 197101, Russia}
\author{Xiangkun Zeng}
\affiliation{MOE Key Laboratory of Advanced Micro-Structured Materials, Shanghai Frontiers Science Center of Digital Optics, Institute of Precision Optical Engineering, and School of Physics Science and Engineering, Tongji University, Shanghai, China}
\author{Di Huang}
\affiliation{MOE Key Laboratory of Advanced Micro-Structured Materials, Shanghai Frontiers Science Center of Digital Optics, Institute of Precision Optical Engineering, and School of Physics Science and Engineering, Tongji University, Shanghai, China}
\author{Tao Jiang}
\affiliation{MOE Key Laboratory of Advanced Micro-Structured Materials, Shanghai Frontiers Science Center of Digital Optics, Institute of Precision Optical Engineering, and School of Physics Science and Engineering, Tongji University, Shanghai, China}
\author{Ivan V. Iorsh}
\affiliation{School of Physics and Engineering, ITMO University, Saint Petersburg 197101, Russia}
\author{Igor Y. Chestnov}
\email{igor.chestnov@metalab.ifmo.ru}
\affiliation{School of Physics and Engineering, ITMO University, Saint Petersburg 197101, Russia}
\author{Vasily Kravtsov}
\email{vasily.kravtsov@metalab.ifmo.ru}
\affiliation{School of Physics and Engineering, ITMO University, Saint Petersburg 197101, Russia}

\begin{abstract}
Layered van der Waals antiferromagnet CrSBr supports strong light--matter coupling and formation of magnetically tunable exciton-polaritons, yet active magnetic control over polariton propagation direction has remained elusive.
Here, we investigate self-hybridized exciton-polaritons in photonic crystal slabs fabricated from CrSBr flakes and their evolution across the antiferromagnetic-to-ferromagnetic spin-flip transition induced by moderate in-plane magnetic fields.
Using angle-resolved reflectance and photoluminescence spectroscopy supported by modeling, we show that the polariton energy continuously tracks the layer-by-layer magnetization switching, revealing a gradual redistribution of oscillator strength from antiferromagnetic to ferromagnetic excitons near the critical field.
Most notably, we demonstrate that the sign of the polariton group velocity can be reversed by a small change in the external magnetic field of only $\sim 40$~mT, resulting in complete switching of the polariton propagation direction.
Our results establish CrSBr photonic crystal slabs as a platform for magnetically controlled polariton transport, opening opportunities for active integrated photonic and polaritonic devices.
\end{abstract}

\maketitle

\noindent
Layered van der Waals semiconductors provide a promising material platform for integrated photonics~\cite{Meng2023, Zotev2025}.
They are characterized by high refractive index, low losses, and strong anisotropy in the visible and near-infrared spectral ranges, while offering straightforward integration with other materials via dry transfer~\cite{Vyshnevyy2023, Ren2025}.
However, tunability of their refractive index and associated optical response~\cite{Gan2022}, which is required for developing active integrated photonic devices like modulators and switches, is usually limited.

Recently, magnetic semiconductor CrSBr has emerged as a unique air-stable 2D material with optical response that can be substantially modified by applying moderate external magnetic fields~\cite{Wilson2021}.
Below the N\'eel temperature of $\sim 132$~K, bulk CrSBr is a layered van der Waals antiferromagnet (AFM) with high and anisotropic refractive index, which supports a strong exciton resonance around 1.36 eV coupled to the magnetic order~\cite{Ziebel2024, Gish2024, Smolenski2025}.
Under external magnetic field, CrSBr can transition into a ferromagnetic (FM) phase, with significant spectral shift of its exciton resonance and corresponding change of the refractive index~\cite{Demir2025, Krelle2025, Datta2025, Smiertka2026}.

Crucially, the high oscillator strength of excitons in CrSBr allows achieving strong and even ultrastrong light--matter coupling with exciton-polariton formation both inside a microcavity and in a self-hybridized geometry~\cite{Dirnberger2023, Wang2023, Li2024afm, Nessi2024, Ziegler2025}.
This provides a powerful approach to controlling optical fields in CrSBr samples through the excitonic fraction of the polaritons~\cite{Li2024arxiv}, leading to magnetically tunable long-range propagation~\cite{Adak2025} or condensation~\cite{Han2025arxiv, Zhang2025arxiv} of polaritons.
Further control of the polaritonic properties in CrSBr, such as polarization, group velocity, and effective mass, can be achieved by engineering the corresponding dispersion via metasurfaces~\cite{Li2024prl}.
Such approaches are well developed for nonmagnetic 2D semiconductors~\cite{Munkhbat2020, Weber2023}, whereas systematic studies of polaritonic states in photonic-crystal architectures based on CrSBr are limited, partially due to the lack of established methods for non-destructive nanostructuring of 2D magnetic flakes.
While in-plane magnetic field tunability of the optical response has been recently demonstrated in nanopatterned CrSBr~\cite{Demir2025}, the associated polariton dispersion evolution remains to be investigated.

Here we study the optical response of CrSBr photonic crystal slabs fabricated by mechanical scanning probe lithography.
Via angle-resolved photoluminescence (PL) and reflectance spectroscopy, we measure the dispersion of self-hybridized exciton-polaritons and reveal its evolution in the spin-flip transition under in-plane magnetic field.
We show that, while spin-flip transitions inside individual CrSBr layers happen in a discrete fashion, the exciton-polariton energy exhibits continuous and sensitive tunability with magnetic field.
Furthermore, we demonstrate a complete switching of exciton-polariton propagation direction with a small change in the external magnetic field of $\sim 40$~mT.
Beyond switching, the broken time-reversal symmetry under an in-plane magnetic field raises the possibility of non-reciprocal polariton propagation, a phenomenon of current interest for topological photonics.
Our results suggest CrSBr as a prospective material platform for developing active integrated photonic and polaritonic devices.

\begin{figure*}[t]
	\includegraphics[width=1\textwidth]{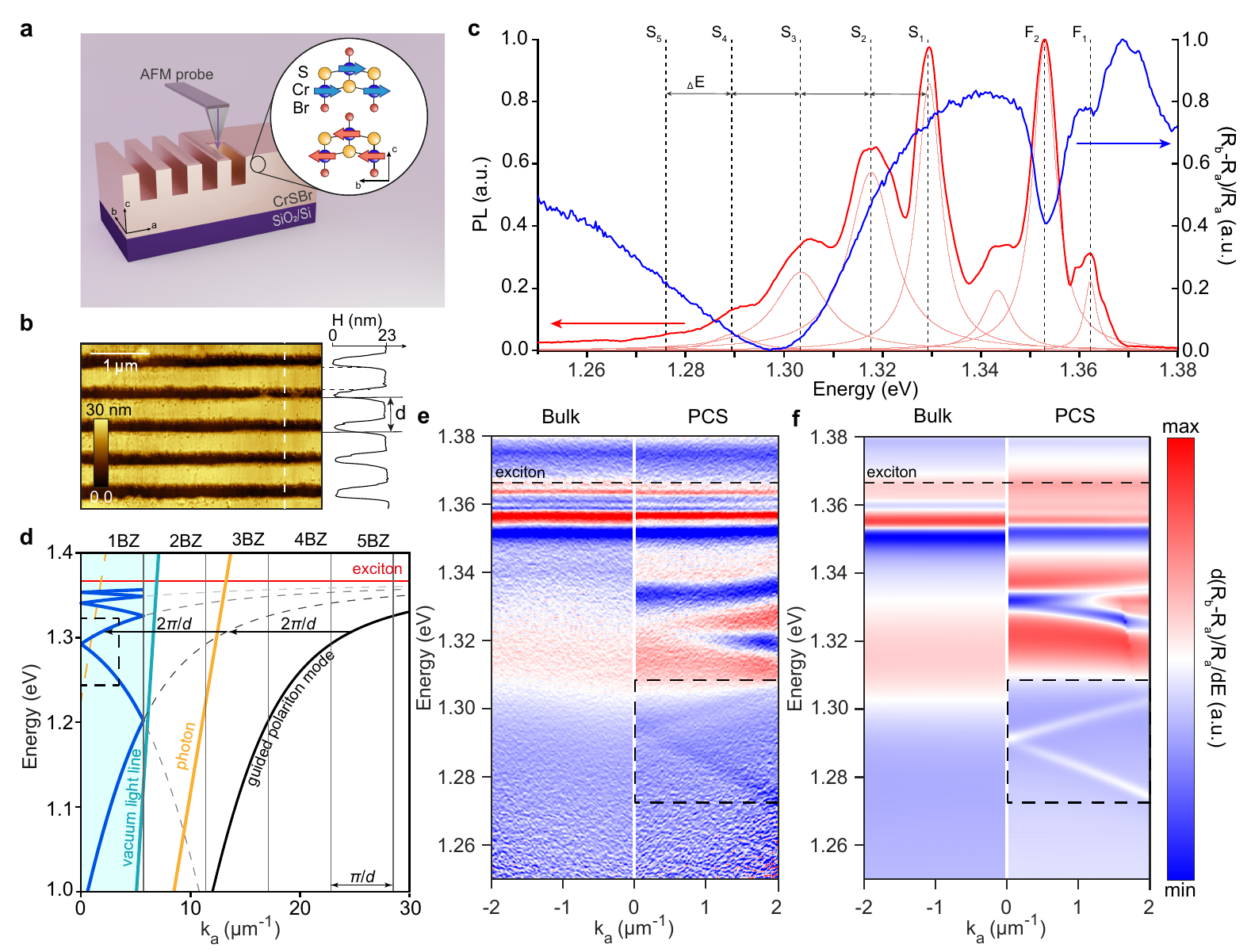}
	\caption{\textbf{Self-hybridized exciton-polaritons in a CrSBr PCS.} 
	(a) Illustration of CrSBr nanostructuring process via scanning probe lithography, with inset showing the crystal structure of CrSBr and spin orientations in the AFM phase.
    (b) Topography of nanopatterned CrSBr measured with atomic force microscopy.
    (c) PL (red curve, left axis) and TE/TM reflectance contrast (blue curve, right axis) spectra measured on an unpatterned 105-nm-thick CrSBr sample, with results of PL fitting by a set of Lorentzian peaks (pink curves).
    (d) Calculated dispersion for CrSBr PCS: a guided polariton mode (black curve) arises from anticrossing between the photonic TE$_0$ mode (yellow curve) and exciton resonance (horizontal red line). The polariton dispersion is folded into the first Brillouin zone (BZ) resulting in photonic-crystal polaritons (blue curve) above the light line (turquoise line).
    (e) First derivative of measured angle-resolved reflectance contrast spectra for unpatterned (left) and patterned (right) CrSBr, with exciton resonance at $1.367$~eV indicated with the horizontal dashed line.
    (f) Corresponding results of numerical simulations for unpatterned (left) and patterned (right) CrSBr films. Black dashed rectangles in (d)-(f) highlight the region with photonic-crystal polariton modes with high group velocities and small wavevectors.}
	\label{fig:Sample}
\end{figure*}

\begin{figure*}[t]
	\includegraphics[width=1\textwidth]{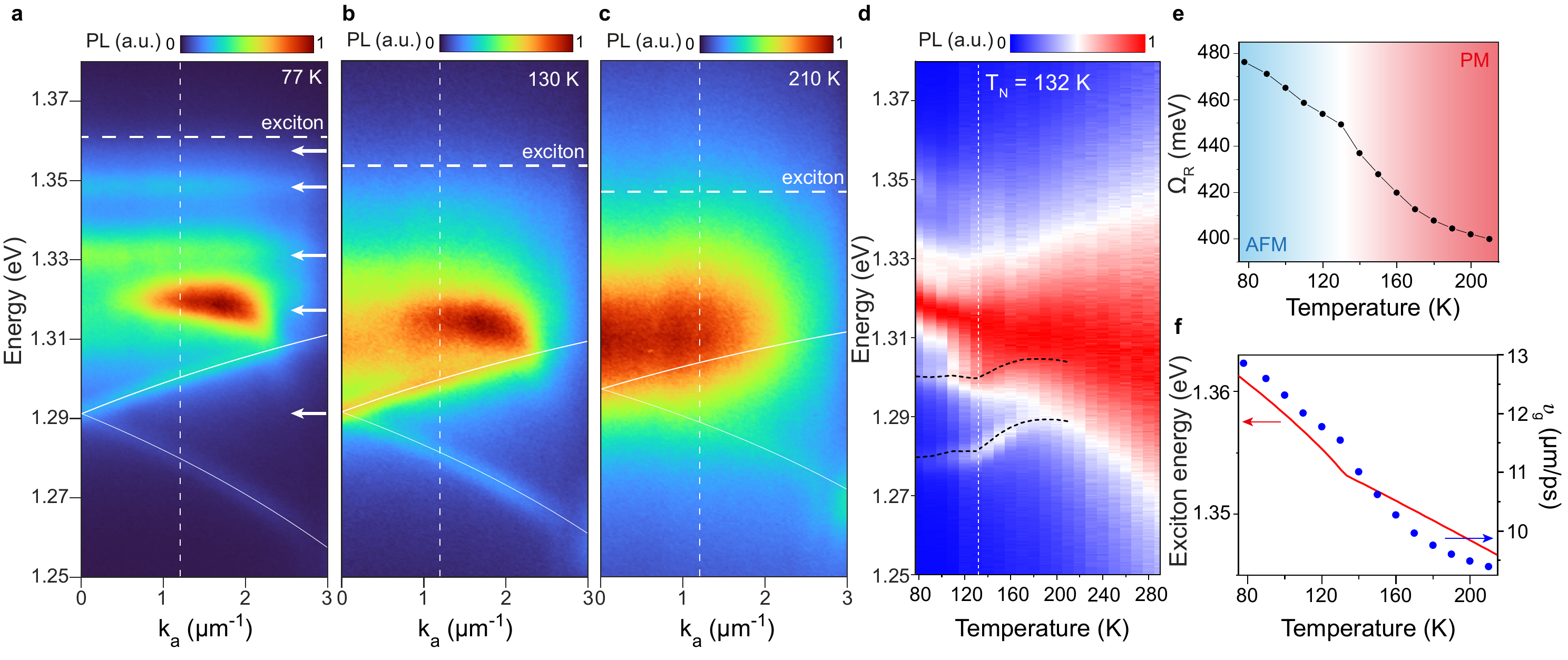}
	\caption{\textbf{Temperature dependence of excitons and polaritons in CrSBr PCS.} 
		(a)-(c) Angle-resolved PL spectra measured on CrSBr PCS at temperatures of (a) 77\,K, (b) 130\,K, and (c) 210\,K, together with exciton energies (horizontal dashed lines) and photonic-crystal polariton dispersions (solid curves) obtained by fitting data to the coupled oscillators model. White arrows indicate peak positions in the PL spectra at $k_a = 0~\mu\mathrm{m}^{-1}$.
		(d) Temperature-dependent PL spectra taken at $k_a = 1.2~\mu\mathrm{m}^{-1}$ as marked in (a)-(c) with vertical dashed lines. The evolution of the polariton energies are shown with black dashed curves, and the vertical dashed line marks the N\'eel temperature $T_N = 132$~K.
		(e) Temperature dependence of the Rabi splitting energy.
		(f) Temperature dependence of the exciton resonance energy (red curve, left axis) and polariton group velocity (blue dots, right axis).}
	\label{fig:Temperature}
\end{figure*}

Photonic crystal slabs (PCS) are fabricated in our experiment from 105-nm-thick CrSBr flakes with mechanical scanning probe lithography using a diamond tip as schematically illustrated in Fig.~\ref{fig:Sample}a and described in Methods (also see Supplementary Note 1).
This etching-free nano-patterning technique provides a non-invasive and low-cost alternative to traditional nanofabrication approaches as previously demonstrated for 2D van der Waals semiconductors and perovskites~\cite{Glebov2023, Alekseev2025}.
Typical surface morphology of a nano-patterned CrSBr flake is shown in an atomic force microscopy  image in Fig.~\ref{fig:Sample}b.

As shown in Fig.~\ref{fig:Sample}c and further discussed in Supplementary Note 2, PL (red) and reflectance (blue) spectra for a non-patterned CrSBr flake measured at 10~K exhibit a multi-peak structure, in general agreement with previously reported results~\cite{Lin2024, Dirnberger2023}.
We attribute the observed higher-energy PL peaks in the $1.347-1.366$~eV range (marked as F$_{1, 2}$) to self-hybridized Fabry--Perot (FP) polariton resonances.
These resonances also appear as Fano-shaped features in the reflectance contrast spectrum calculated as $(R_\mathrm{b} - R_\mathrm{a})/R_\mathrm{a}$, where $R_\mathrm{b, a}$ are signals reflected in the TE ($E$-field is along the $b$-axis) and TM ($E$-field is along the $a$-axis) polarizations, respectively. 
Numerical modeling (see Supplementary Note 3) shows that the observed spectral features can be described by considering interaction of a single exciton resonance at $1.36$~eV with FP modes within the crystal enabled by the drastic modulation of dielectric permittivity in the vicinity of the exciton resonance~\cite{Dirnberger2023}.
Additionally, the PL spectrum shows several lower-energy peaks in the range $1.265-1.336$~eV with possible contributions from phonon replicas and surface exciton states~\cite{Lin2024, Shao2025}, which are weakly coupled to light and do not appear in the reflectance spectrum. 

Due to its high refractive index, the CrSBr flake supports a TE waveguide mode as shown by calculation results plotted in Fig.~\ref{fig:Sample}d (yellow curve), which strongly couples to the exciton resonance forming a guided exciton-polariton mode (black) deep below the light line (turquoise).
Periodic modulation in the CrSBr PCS folds the polariton mode inside the first Brillouin zone resulting in self-hybridized photonic-crystal exciton-polaritons with dispersion (blue curve) that lies above the light line and efficiently couples to the far field.

We obtain the polariton dispersion in experiment via angle-resolved reflectance measurements~\cite{Kravtsov2020, Khestanova2024} as outlined in Methods (also see Supplementary Note 4), with the corresponding wavevector dependencies of the TE/TM reflectance contrast first derivative with respect to energy plotted in Fig.~\ref{fig:Sample}e for the unpatterned flake (left) and PCS (right).
While the data from the unpatterned CrSBr flake shows only almost dispersionless self-hybridized FP polariton modes, the patterned CrSBr in addition exhibits highly dispersive exciton-polariton branches ($1.27-1.31$~eV, dashed rectangle) characterized by high values of group velocity reaching $1.3\times10^7$~m/s.
We identify these dispersion branches as lower polariton states formed due to strong coupling between the TE$_0$ optical mode and CrSBr exciton resonance (see Supplementary Note 5 for data measured in PCS samples with different parameters).
Additionally, lower polariton states arising from exciton coupling with the TE$_1$ optical mode are observed at higher energies of $1.32-1.34$~eV.

The observed experimental data are in good agreement with the results of numerical simulations carried out using the Fourier modal method as shown in Fig.~\ref{fig:Sample}f.
The excitonic response is considered as a Lorentz permittivity function along the $b$-axis: $    \varepsilon_{bb}(E) = \varepsilon_{bg} + f/\left(E_{\rm ex}^2-E^2-i\Gamma E\right)$, where $\varepsilon_{bg}$ is a background permittivity constant, $E_{\rm ex}$, $\Gamma$, and $f$ are the exciton resonance energy, damping rate, and oscillator strength, respectively.
The best match with the experimental data is achieved with $E_{\rm ex} = 1.367$~eV, $\Gamma = 6$~meV, and $f=2.5\,{\rm{eV}}^2$, consistent with values quoted in previous reports~\cite{Wang2023,Dirnberger2023,Li2024afm}.

The assignment of exciton–polariton branches in the reflectance spectra is corroborated with temperature-dependent angle-resolved PL measurements shown in Fig.~\ref{fig:Temperature}.
PL spectra taken at 77 K (a), which is well below the N\'eel temperature of $\mathrm{T_N} = 132$~K, exhibit a series of characteristic features marked with white arrows.
Similar to the case of unpatterned CrSBr (Fig.~\ref{fig:Sample}d) yet with slightly different energies due to the elevated temperature, higher-energy peaks corresponding to self-hybridized FP polaritons are resolved at 1.348~eV and 1.357~eV, and emission from weakly coupled lower-energy excitonic peaks is observed in the $1.30-1.34$~eV range.
In accordance with angle-resolved reflectance data (Fig.~\ref{fig:Sample}e), the lower polariton branches of two distinct modes are observed in the angle-resolved PL spectra: TE$_0$ in the range of $1.264–1.308$~eV and TE$_1$ in the range of $1.308–1.324$~eV.
The dispersion of the TE$_0$ polaritons is well described within the coupled oscillator model (see Supplementary Note 1) as plotted with solid white curves in Fig.~\ref{fig:Temperature}a, yielding a Rabi splitting energy of $\hbar\Omega_\mathrm{R} = 476$~meV at 77~K.

Increasing the sample temperature leads to substantial modification of the angle-resolved PL spectra.
As illustrated in Fig.~\ref{fig:Temperature}b,c for 130 K and 210 K (see more data in Supplementary Note 6), the excitonic peaks redshift and broaden, whereas the polariton branches blueshift and eventually become indistinguishable.
To provide a comprehensive visualization of these trends, Fig.~\ref{fig:Temperature}d displays the full temperature dependence of the PL spectrum at a fixed in-plane wavevector $k_a = 1.2~\mu\mathrm{m}^{-1}$ marked in panels (a)-(c) with vertical dashed lines.
Notably, while strong exciton--photon coupling persists in the paramagnetic (PM) phase up to at least $\sim 200$~K where polariton branches are still discernible, the temperature dependencies of the polariton energies in Fig.~\ref{fig:Temperature}d (dashed black curves) exhibit a distinct change of slope in the vicinity of the Néel temperature $T_N$ (vertical dashed line).

This behavior can be understood by considering the corresponding temperature dependencies of the fitted Rabi splitting $\Omega_\mathrm{R}$ and exciton energy $E_\mathrm{ex}$ shown in Fig.~\ref{fig:Temperature}e,f.
Below $T_N$, the decreasing trends of both $\Omega_\mathrm{R}$ and $E_\mathrm{ex}$ compensate each other keeping the polariton energies almost independent of temperature.
Above $T_N$, in the absence of magnon-related spectral shifts (see Supplementary Note 7), $E_\mathrm{ex}$ exhibits a slower decrease with temperature (f).
In contrast, $\Omega_\mathrm{R}$ starts decreasing faster (e), which indicates a reduction in exciton oscillator strength $f$ under the AFM--PM transition, ultimately resulting in the rapid blueshift of polariton energies observed for $T > T_N$ in Fig.~\ref{fig:Temperature}d.
Interestingly, polariton modes also exhibit a decrease in group velocity with increasing temperature as shown in Fig.~\ref{fig:Temperature}f (blue dots, right axis), which potentially enables thermal control of polariton propagation speed in CrSBr waveguides and metasurfaces.

\begin{figure*}[t]
	\includegraphics[width=1\textwidth]{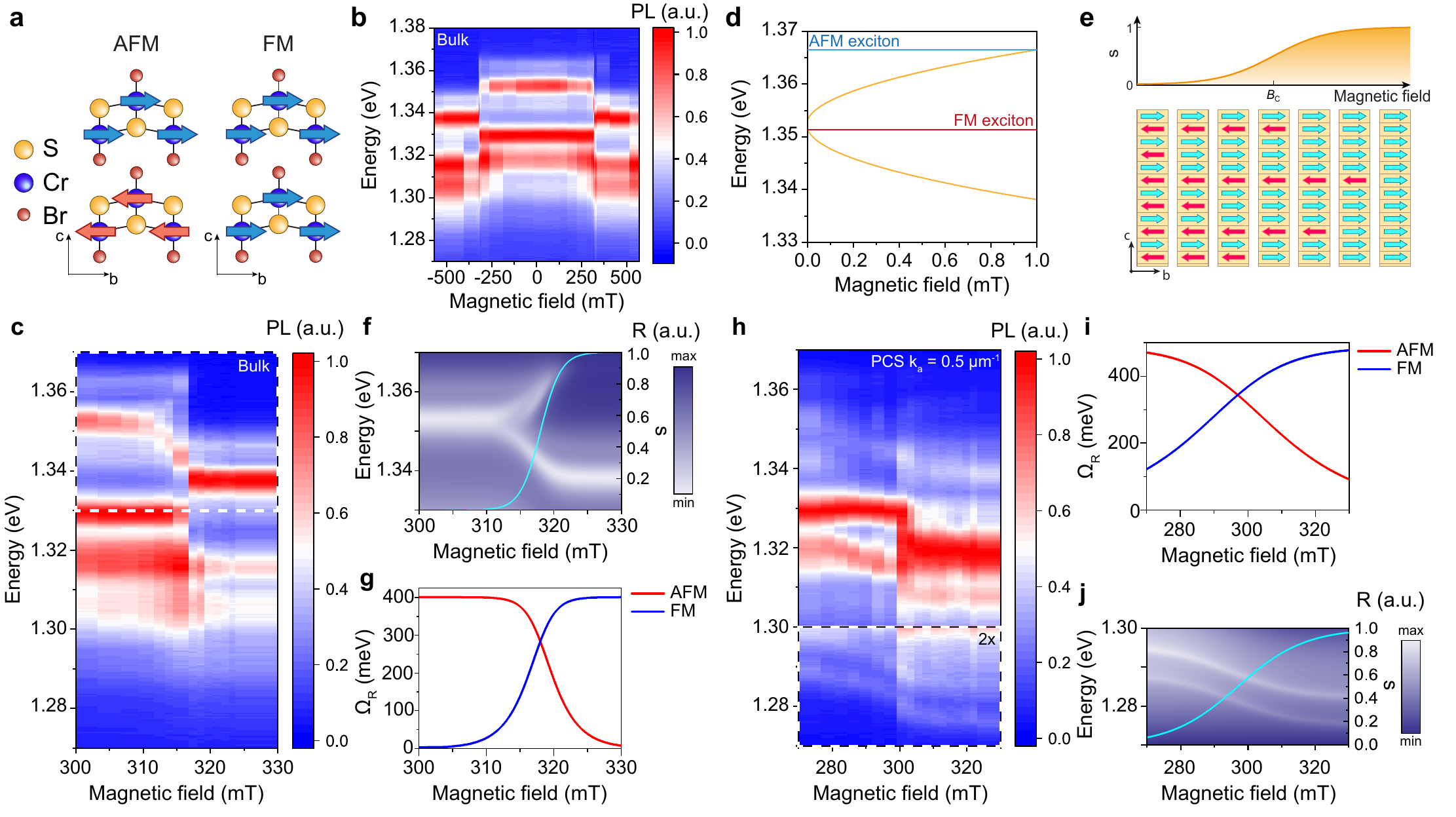}
	\caption{\textbf{Magnetic-field control of FP and photonic-crystal polaritons in CrSBr.} 
		(a) Spin orientations in the CrSBr crystal lattice (side view) in the AFM (left) and FM (right) phases for the spin-flip transition under external magnetic field $B$ parallel to the b-axis.
		(b) Magnetic field dependence of normalized PL spectra measured on the unpatterned CrSBr sample.
		(c) Corresponding magnetic field dependence near the spin-flip transition.
		(d) Calculated via a 3 coupled oscillators model exciton-polariton energies (orange curves) near the AFM--FM transition as functions of parameter $s$ describing the density ratio of AFM (blue line) and FM (red line) excitons.
		(e) Magnetic-field dependence of the fraction of FM excitons (top), illustration of the layer-by-layer AFM to FM switching in the CrSBr sample (bottom).
		(f) Simulated magnetic field dependencies of the reflectance spectrum for the unpatterned CrSBr near the AFM--FM transition (image plot) and parameter $s$ (cyan curve).
		(g) Calculated Rabi splittings for the FP mode coupled to AFM (red curve) and FM (blue curve) excitons.
		(h) Magnetic field dependence of normalized PL spectra measured on the CrSBr PCS sample across the AFM–-FM transition region at $k_a = 0.5~\micron^{-1}$.
		(i) Calculated Rabi splittings for the photonic-crystal mode coupled to AFM (red curve) and FM (blue curve) excitons.
		(j) Simulated magnetic field dependencies of the reflectance spectrum for the CrSBr PCS sample near the AFM--FM transition (image plot) and parameter $s$ (cyan curve).}
	\label{fig:Magnetic}
\end{figure*}

\begin{figure*}[t]
	\includegraphics[width=1\textwidth]{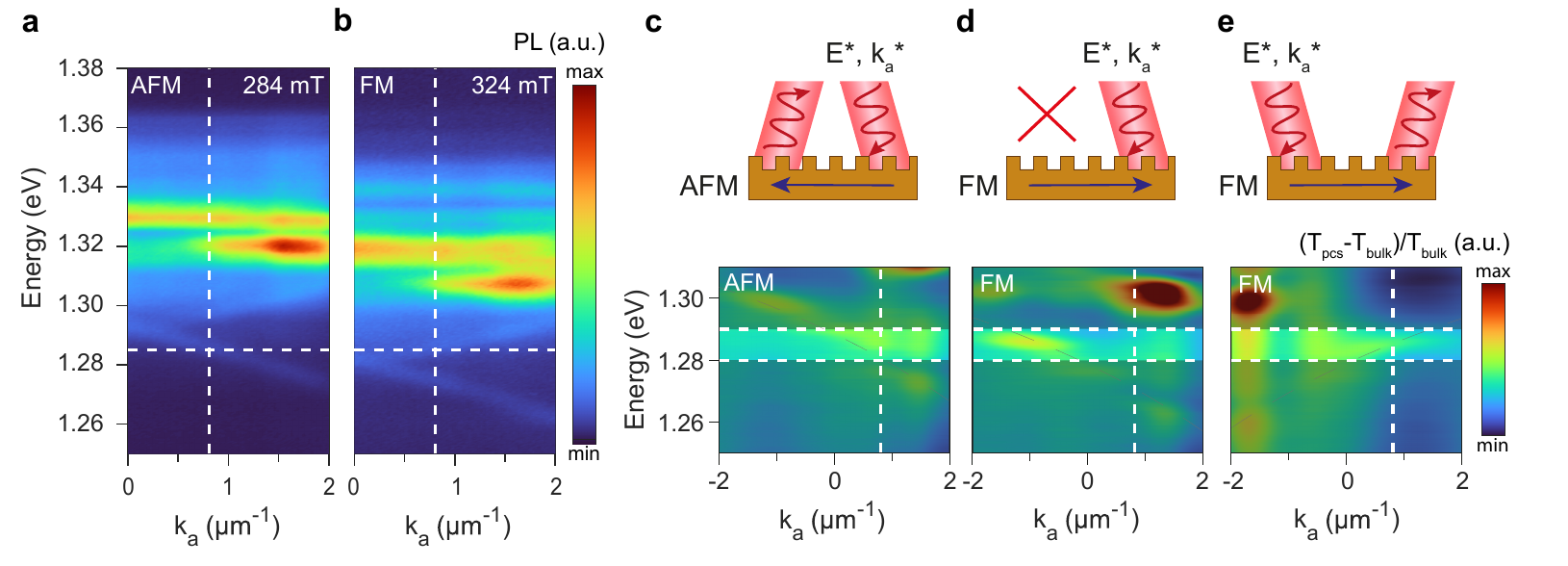}
	\caption{\textbf{Switching of polariton propagation direction.}
    (a, b) Angle-resolved PL spectra of the CrSBr PCS in the AFM and FM phases. White dashed lines indicate the energy $E^*$ and wave vector $k^*_a$ corresponding to switching sign of the polariton group velocity.
    (c-e) Schematic side-view illustrations of three transmission measurements (top), with optical excitation/detection indicated with red wavy arrows and polariton propagation direction indicated with blue straight arrow. 
    Corresponding measured differential transmission signals (bottom), with black curves indicating the propagating polariton dispersion branches and vertical dashed lines marking the probed wave vector $k^*_a$. The spectral region of interest near $E^*$ is indicated with a dashed rectangle.}
	\label{fig:Switching}
\end{figure*}

Having established the origin and parameters of the polariton response in our CrSBr PCS, we next study its modification under an external magnetic field.
CrSBr is known to exhibit an AFM--FM transition driven by a spin-flip process along the easy magnetization axis~\cite{Wilson2021} as schematically illustrated in Fig.~\ref{fig:Magnetic}a.
This transition activates previously forbidden interlayer hybridization of electron and hole orbitals, resulting in a jump-like modification of the exciton response~\cite{Shi2025}.
Applying a varying magnetic field $B$ parallel to the $b$-axis of the unpatterned CrSBr flake, we observe such transition in the PL spectra shown in Fig.~\ref{fig:Magnetic}b, with a distinct $\sim 15$~meV redshift of PL peaks for $B \gtrsim 315$~mT.

A detailed examination of the magnetic phase transition region (Fig.~\ref{fig:Magnetic}c) reveals a more complex dependence of the PL spectrum on magnetic field: the self-hybridized FP polaritonic peak at $1.352$~eV (spectral region indicated with dashed rectangle) undergoes a Y-shaped splitting into 2 distinct branches followed by their gradual shifting in opposite directions (compare to calculation results shown in Fig.~\ref{fig:Magnetic}f).
To understand this unusual behavior, we consider a model where AFM and FM phases coexist in the sample giving rise to 2 distinct excitonic states with different resonant energies, which both strongly couple to the same optical mode (see Supplementary Note 8).
As a result, 2 hybrid polariton states emerge with energies plotted in Fig.~\ref{fig:Magnetic}d (orange curves) as functions of the parameter $s$ describing the relative density of FM and AFM excitons.

Since our sample contains a large number of layers, switching between AFM and FM phases can occur layer-by-layer in a multi-step process characteristic of A-type antiferromagnets~\cite{Sun2025, Wang2026} where the relative density of AFM and FM excitons $s$ gradually and nonlinearly changes with the applied magnetic field, as schematically illustrated in Fig.~\ref{fig:Magnetic}e.
To assess the evolution of $s$ with magnetic field, we find the best match between $B$-dependent calculated and measured polariton spectra, with the corresponding results of calculations and obtained $s(B)$ dependence plotted in Fig.~\ref{fig:Magnetic}f as the image plot and curve, respectively.
Our analysis shows that the magnetization switching process in the unpatterned CrSBr sample can be described with the following function: $s(B) = \left(1 + \tanh\left((B - B_c)/\Delta B\right) \right)/2$ with critical field $B_c = 318$~mT and transition width $\Delta B = 3$~mT.
The Rabi splitting values for light interacting with AFM and FM excitons follow the same trend revealing a gradual redistribution of the effective oscillator strength from AFM to FM excitons as shown in Fig.~\ref{fig:Magnetic}g.

Similar magnetic-field-dependent behavior is observed for photonic-crystal polaritons in the patterned sample.
Fig.~\ref{fig:Magnetic}h shows the evolution of PL spectra near the spin-flip transition measured for the CrSBr PCS at a selected in-plane wavevector $k_a = 0.5~\micron^{-1}$ (see Supplementary Note 9 for other wavevectors).
As seen in the plot (dashed rectangular region), both photonic-crystal polariton peaks at $1.289$~eV and $1.299$~eV exhibit a continuous redshift, which again can be described using the model of 3 coupled oscillators including AFM excitons, FM excitons, and the guided optical mode as detailed in Supplementary Note 8.
Matching the calculated and experimental $B$-dependencies of polariton energies, we obtain the evolution of Rabi splitting values for AFM and FM excitons (i), polaritonic spectra (j, image plot), and parameter $s$ (j, curve) with magnetic field.
The corresponding values of critical field $B_c = 297$~mT and transition width $\Delta B = 20$~mT are somewhat different than those obtained for unpatterned CrSBr.
We attribute this difference to structural modification of the sample during photonic crystal fabrication, resulting in both a lower effective thickness and its stronger local variation within the probed area, with correspondingly altered layer-by-layer magnetic switching behavior.

Finally, we discuss the change in group velocity of photonic-crystal polaritons under the external magnetic field.
Fig.~\ref{fig:Switching}a,b shows angle-resolved PL spectra measured on the patterned CrSBr sample in the AFM (a, $B = 284$~mT) and FM (b, $B = 324$~mT) phases.
As seen in the plots, for selected values of energy $E^* = 1.283$~eV (horizontal dashed line) and in-plane wavevector $k^*_a = 0.81$~\micron$^{-1}$ (vertical dashed line), exciton-polaritons completely switch their propagation direction upon the AFM--FM transition.
Indeed, in the AFM phase (a), the corresponding polariton group velocity is extracted from the dispersion as $v^\mathrm{AFM}_g = -13.4$~\micron/ps, while in the FM phase it changes to $v^\mathrm{FM}_g = +13.0$~\micron/ps.
The observed behavior is enabled in our CrSBr PCS sample by the existence of 2 distinct but spectrally close polariton dispersion branches with opposite slopes, in combination with the sizable overall redshift of polariton resonances upon the AFM--FM transition. 

In order to test this polariton propagation direction switching effect, we carry out the following measurement.
Using spatial filtering, we excite one half of the PCS sample and detect the emission of transmitted polaritons from the other half as schematically illustrated in Fig.~\ref{fig:Switching}c-e (top panels).
When we excite the right side of the sample in the AFM phase (red wavy arrow in Fig.~\ref{fig:Switching}c, top panel), polaritons with negative group velocity propagate from right to left (blue arrow) and scatter to far-field, which is evidenced by the enhanced signal registered at $k^*_a = 0.81$~\micron$^{-1}$ and $E = 1.283$~eV (Fig.~\ref{fig:Switching}c, bottom panel).
After switching to the FM phase, polaritons acquire positive group velocity and propagate in the opposite direction, which results in a significantly reduced signal at the selected energy and wavevector (Fig.~\ref{fig:Switching}d, bottom panel).
If we instead excite the left side of the sample in the FM phase and collect emission from the right side, the signal recovers (Fig.~\ref{fig:Switching}d, bottom panel), which indicates that the polaritons are indeed propagating from left to right in this case.

In summary, our work demonstrates that self-hybridized exciton-polaritons with large group velocities can be formed in photonic crystal slabs made from thin CrSBr layers via non-destructive scanning probe lithography.
Signatures of these photonic-crystal polaritons with Rabi splitting $> 450$~meV are observed in the dispersion for temperatures up to $\sim 200$~K.
In moderate external magnetic fields of $\sim 0.3$~T oriented along the $b$-axis, when CrSBr undergoes a spin-flip transition from AFM to FM phase, exciton and polariton resonances spectrally redshift by $\sim 15$~meV.
In contrast to an abrupt redshift of the exciton resonance, both FP and photonic-crystal polaritons exhibit a much smoother redshift, since they are simultaneously coupled to both AFM and FM excitons coexisting in the sample near the critical field $B_c$.
This behavior allows us to reveal the exact evolution of the AFM to FM exciton density ratio during the spin-flip transition. 
Additionally, we show that switching between the AFM and FM phases in CrSBr photonic crystals via changing the external field by only $\sim 40$~mT can result in changing the sign of polariton group velocity, effectively switching the propagation direction of polaritons at a certain energy and wavevector.
Our results suggest that self-hybridized photonic-crystal exciton-polaritons in patterned CrSBr flakes provide a unique platform for developing magnetically-controlled integrated photonic circuits and potentially for studying time-reversal-symmetry-broken non-reciprocal polariton propagation.\\


\noindent
{\bf \large Methods}\\
\noindent
\textbf{Sample fabrication.}
Thin CrSBr flakes were obtained from bulk CrSBr crystals by mechanical exfoliation onto polydimethylsiloxane (PDMS) films with Nitto adhesive tape.
Large uniform flakes were identified in an optical microscope and dry-transferred from PDMS onto SiO$_2$/Si substrates with a \SI{1}{\micro\meter}-thick oxide layer using a home-built transfer setup.
Photonic crystals were fabricated in CrSBr films by engraving one-dimensional gratings employing mechanical scanning probe nanolithography with single-crystal diamond tip cantilevers DRP\_In (TipsNano) with $100-600$~N/m force constant operated in the contact mode.
Grooves were oriented along the crystallographic \textit{b} axis.
After patterning, the topography was characterized by atomic force microscopy in the tapping mode. The structure investigated in this work had a period of 551 nm, a modulation depth of 23 nm, and a fill factor of 0.64 and was formed on a bulk crystal with a thickness of 105 nm.

\noindent
\textbf{Optical measurements}
Temperature-dependent measurements were carried out in a liquid-nitrogen flow cryostat (Linkam Scientific) at sample temperature varied from $77$~K to $300$~K.
Magnetic-field-dependent measurements were performed in an ultra low vibration closed cycle helium cryostat (Advanced Research Systems) at \SI{10}{\kelvin}, with magnetic field applied parallel to the crystallographic \textit{b} axis in CrSBr crystals.
A He--Ne laser and broadband halogen lamp were used as excitation sources for PL and reflectance measurements, respectively.
Excitation and collection were performed using a long-working-distance microscope objective (Mitutoyo NIR HR 50$\times$, $0.65$).
Angle-resolved reflectance and PL spectra were recorded in the back-focal plane of the objective using a slit spectrometer (Princeton Instruments SP2500) equipped with a 150~lines/mm diffraction grating and a liquid-nitrogen-cooled CCD camera.
Polarization and wavevector directions were selected in the detection optical path using a Glan--Taylor polarizer, half-wave plate, and Dove prism.\\

\noindent
{\bf \large Acknowledgments}\\
\noindent
Sample fabrication was supported by Priority 2030 Federal Academic Leadership Program.
Optical measurements were supported by Russian Science Foundation project 25-72-20029.\\

\noindent
{\bf \large References}
\bibliography{crsbrgratings}  
\bibliographystyle{naturemag}

\end{document}